\documentclass[sn-mathphys-num]{sn-jnl}

\usepackage{graphicx}%
\usepackage{multirow}%
\usepackage{amsmath,amssymb,amsfonts}%
\usepackage{amsthm}%
\usepackage{mathrsfs}%
\usepackage[title]{appendix}%
\usepackage{xcolor}%
\usepackage{textcomp}%
\usepackage{manyfoot}%
\usepackage{booktabs}%
\usepackage{algorithm}%
\usepackage{algorithmicx}%
\usepackage{algpseudocode}%
\usepackage{listings}%

\usepackage{physics}
\usepackage[T1]{fontenc}
\usepackage[figurename=Figure]{caption}
\usepackage{float}    
\usepackage{circuitikz}
\usepackage{enumitem} 
\usepackage{listings} 
\usepackage{tikz}
\usepackage[makeroom]{cancel}

\usepackage{tcolorbox}
\usepackage{array}

\usepackage[font=footnotesize,labelfont=bf]{caption}
\usepackage{subcaption}

\usetikzlibrary{positioning}
\usetikzlibrary{angles}
\usepackage{listofitems} 
\tikzstyle{mynode}=[thick,draw=blue,fill=blue!20,circle,minimum size=22]

\usepackage[]{mdframed}
\usepackage{multicol}
\usepackage{pstricks}

\usetikzlibrary{arrows.meta} 
\usepackage[outline]{contour} 
\contourlength{1.4pt}
\tikzset{>=latex} 
\usepackage{xcolor}
\colorlet{myred}{red!80!black}
\colorlet{myblue}{blue!80!black}
\colorlet{mygreen}{green!60!black}
\colorlet{myorange}{orange!70!red!60!black}
\colorlet{mydarkred}{red!30!black}
\colorlet{mydarkblue}{blue!40!black}
\colorlet{mydarkgreen}{green!30!black}
\tikzstyle{node}=[thick,circle,draw=myblue,minimum size=22,inner sep=0.5,outer sep=0.6]
\tikzstyle{node in}=[node,green!20!black,draw=mygreen!30!black,fill=mygreen!25]
\tikzstyle{node hidden}=[node,blue!20!black,draw=myblue!30!black,fill=myblue!20]
\tikzstyle{node convol}=[node,orange!20!black,draw=myorange!30!black,fill=myorange!20]
\tikzstyle{node out}=[node,red!20!black,draw=myred!30!black,fill=myred!20]
\tikzstyle{connect}=[thick,mydarkblue] 
\tikzstyle{connect arrow}=[-{Latex[length=4,width=3.5]},thick,mydarkblue,shorten <=0.5,shorten >=1]
\tikzset{ 
  node 1/.style={node in},
  node 2/.style={node hidden},
  node 3/.style={node out},
}
\def\nstyle{int(\lay<\Nnodlen?min(2,\lay):3)} 

\usetikzlibrary{shapes,arrows,calc,positioning,fit} 

\tikzset{
   block/.style = {draw, rectangle,
       minimum height=1cm,
       align = center
   },
   input/.style = {coordinate,node distance=1cm},
   output/.style = {coordinate,node distance=1cm},
   arrow/.style={draw, -latex,node distance=2cm},
   pinstyle/.style = {pin edge={latex-, black,node distance=2cm}},
   sum/.style = {draw, circle, node distance=1cm},
   gain/.style = {
     regular polygon, regular polygon sides=3,
     draw, fill=white, text width=1em,
     inner sep=0mm, outer sep=0mm,
     shape border rotate=-90
   },
   dot/.style={circle,fill,draw,inner sep=0pt,minimum size=3pt}
 }

\newcommand{\spinUp}{\ket{\psi^{\uparrow}}}
\newcommand{\spinDown}{\ket{\psi^{\downarrow}}}
\newcommand{\spinUpUp}{\ket{\psi^{\uparrow \uparrow}}}
\newcommand{\spinDownDown}{\ket{\psi^{\downarrow \downarrow}}}

\newcommand{\spinDownUpPlus}{\ket{\psi^{\downarrow \uparrow}_+}}
\newcommand{\spinDownUpMinus}{\ket{\psi^{\downarrow \uparrow}_-}}

\theoremstyle{thmstyleone}%
\theoremstyle{thmstyletwo}%

\theoremstyle{thmstylethree}%

\begin{document}

\title[Fast generation of entanglement between coupled spins using optimization and deep learning methods]{Fast generation of entanglement between coupled spins using optimization and deep learning methods}


\newcommand{\orcidauthorA}{0009-0002-7949-5087} 
\newcommand{\orcidauthorB}{0000-0002-3945-4304} 
\newcommand{\orcidauthorC}{0000-0001-5206-2244} 







\author[1]{\fnm{Dimitris} \sur{Koutromanos}}\email{dkoutromanos@upatras.gr}

\author[1]{\fnm{Dionisis} \sur{Stefanatos}}\email{dstefanatos@upatras.gr}

\author[1]{\fnm{Emmanuel} \sur{Paspalakis}}\email{paspalak@upatras.gr}

\affil[1]{\orgdiv{Materials Science Department}, \orgname{School of Natural Sciences, University of Patras}, \postcode{26504}, \city{Patras}, \country{Greece}}


\abstract{Coupled spins form composite quantum systems which play an important role in many quantum technology applications, with an essential task often being the efficient generation of entanglement between two constituent qubits. The simplest such system is a pair of spins-$1/2$ coupled with Ising interaction, and in previous works various quantum control methods such as adiabatic processes, shortcuts to adiabaticity and optimal control have been employed to quickly generate there one of the maximally entangled Bell states. In this study, we use machine learning and optimization methods to produce maximally entangled states in minimum time, with the Rabi frequency and the detuning used as bounded control functions. We do not target a specific maximally entangled state, like the preceding studies, but rather find the controls which maximize the concurrence, leading thus automatically the system to the closest such state in shorter time. By increasing the bounds of the control functions we observe that the corresponding optimally selected maximally entangled state also changes and the necessary time to reach it is reduced. The present work demonstrates also that machine learning and optimization offer efficient and flexible techniques for the fast generation of entanglement in coupled spin systems, and we plan to extent it to systems involving more spins, for example spin chains.}




\maketitle

\section{Introduction}

Coupled spin systems play a crucial role in quantum information processing, for example in quantum state transfer \cite{Bose03,Burgarth05}, quantum communication \cite{Murphy10}, and quantum computation \cite{Marx98,Mitrikas08,Burgarth10}. An important task frequently encountered in quantum technology applications involving these systems is the fast generation of entanglement between two component qubits \cite{bayat2022entanglement,Galve09,Estarellas17,Bazhanov18}. The simplest coupled spin system consists of two spins-$1/2$ with Ising coupling between them \cite{unanyan2001preparation,Stefanatos04,paul2016high,stefanatos2019efficient,Chen20,Yang20}. The importance of this prototype system emanates from that it can be used to model realistic quantum systems, for example the exciton-biexciton system in two coupled quantum dots \cite{quiroga1999entangled,kis2004controlled,creatore2012creation,StefanatosPRA2020}. In the present study we concentrate on the problem of fast entanglement generation in this generic quantum system.

Quantum control \cite{dong2010quantum,d2021introduction,shore2011manipulating} is the discipline that tries to control quantum systems and drive them to a desired state or behaviour. In previous works \cite{unanyan2001preparation,paul2016high,zhang2017reverse,yu2018fast,stefanatos2019efficient,Chen20,Yang20}, various quantum control methods have been used to efficiently generate one of the maximally entangled Bell states in our coupled spin pair, when starting from the initial spin-down state. Specifically, adiabatic rapid passage was employed in Ref. \cite{unanyan2001preparation} for implementing the desired population transfer. Adiabatic processes provide robust solutions to the problem but require long time evolution, which might have detrimental effects in the presence of losses. Later on, shortcuts to adiabaticity methods \cite{guery2019shortcuts,stefanatos2021shortcut} were used to speed up the evolution \cite{paul2016high,zhang2017reverse,yu2018fast,stefanatos2019efficient}, by driving the system to the target Bell state without following the slow adiabatic trajectory. Additionally, numerical optimal control
was exploited to generate the desired Bell state in the minimum possible time, under bound constraints on the control fields in Ref. \cite{stefanatos2019efficient} and with unbounded local controls in Refs. \cite{Chen20,Yang20}.

In the present article, building upon our previous work \cite{stefanatos2019efficient}, we do not target a specific maximally entangled state, like the preceding studies, but rather find the controls which maximize the concurrence in the coupled spin pair, driving thus automatically the system to the closest such state in shorter time. We use as control functions the Rabi frequency and the detuning, both bounded,  and solve the corresponding optimal control problem using two optimization methods, Particle Swarm Optimization (PSO) \cite{eberhart1995particle,parsopoulos2002particle} and Sequential Least Squares Programming (SLSQP) \cite{Schittkowski82}, as well as a machine learning and specifically Deep Learning (DL) method \cite{goodfellow2016deep,lecun2015deep}. With each method we find optimal controls which achieve concurrence levels larger than $0.9999$ in minimum time. For comparison, we also use them to find the controls which obtain the same levels of fidelity for the specific Bell state, under the same control bound. We observe that the desired concurrence target is reached faster, while the corresponding maximally entangled state is different than the particular Bell state. By increasing the control bound we notice that the targeted concurrence level is obtained faster, while a different maximally entangled state is reached for each bound.

Regarding the methods used, PSO is an evolutionary algorithm \cite{zhang2015comprehensive}, and such tools have been extensively used for numerical quantum control \cite{Pechen06,Zahedinejad14,Dong20,Mortimer21,Brown22}. PSO has been exploited in the context of quantum control within Lyapunov control for enhanced population transfer between quantum states in a multi-level quantum system \cite{GuanIEEEAccess2020} and more recently in quantum optics for enhancing dynamical photon blockade \cite{Zhang24}. On the other hand, the SLSQP algorithm has been used in combination with reinforcement learning for the fast ground state preparation in a many-body system, inspired by counterdiabatic driving \cite{Yao21}.
Another way to solve the numerical optimal control problem is by utilizing machine learning and specifically DL methods. 
Note that machine learning has already been used to solve quantum control problems \cite{dong2023learning,Wu19,krenn2023artificial}, with a very successful method being that of Reinforcement Learning (RL) \cite{sutton2018reinforcement,szepesvari2022algorithms}, where interaction with the quantum system helps AI agents to develop optimal behaviours and discover optimal pulses \cite{bukov2018reinforcement,niu2019universal,porotti2019coherent,zhang2019does,ZhengEPL2019,paparelle2020digitally,Yao21,ding2021breaking,HeEPJQTech2021,giannelli2022tutorial,ma2022curriculum,LiuSoftComp2022,KoutromanosInfo2024,YuIEEETrArtInt2024,NguyenMLST2024}. 

Lately, a whole new field has been developed, at the intersection of DL and physics, called Physics Informed Machine Learning  \cite{karniadakis2021physics}, which studies and develops methods trying to simulate and control physical systems efficiently. One of the most successful models is the Physics Informed Neural Network (PINN), which is a feed-forward neural network mainly used to predict the evolution of a dynamical system \cite{karniadakis2021physics,raissi2019physics}. This model is extended to the optimal control area \cite{antonelo2024physics} and has also been used for controlling quantum systems \cite{norambuena2024physics}. Here, we propose a novel model based on PINNs but without the extra effort needed to predict the actual dynamics of the system, which is governed by the Schr\"{o}dinger equation. The new model only predicts the control fields that are used to simulate the dynamics of the quantum system and produce the final state, which eventually enters the calculation of the loss (objective) function. We train the model in two settings, one by maximizing the concurrence of the final state and another by maximizing the fidelity of the transfer to the specific Bell state, for comparison.
The present work demonstrates that machine learning and optimization offer efficient and flexible techniques for the fast generation of entanglement in coupled spin systems, and we plan to extent it to systems involving more spins, for example spin chains.

This article starts with the description of the quantum system and the definition of the optimal control problem under study in section \ref{sect:coupledSpins}. The optimization and DL methods used are presented in Section \ref{sect:optim}. The results obtained and the corresponding analysis are presented in Section \ref{sect:results}. The present study concludes with Section \ref{sect:concl}.


\section{Generation of entanglement in a pair of coupled spins as an optimal control problem}
\label{sect:coupledSpins}

In this section we present the system and the problem studied in the paper.

\subsection{A pair of spins-$1/2$ coupled with Ising interaction}

The simplest example of a coupled spin system is a pair of spin-$1/2$ with Ising interaction along the $z$-axis. We additionally consider that both spins interact with the same time-dependent magnetic field $\mathbf{B}(t) = [B_x(t), B_y(t), B_z(t)]$, thus the corresponding Hamiltonian is \cite{unanyan2001preparation,paul2016high,stefanatos2019efficient}: 
\begin{align}
\label{Hamiltonian}
    H(t) = 4 \xi S_{1z} \otimes S_{2z} + \mu \mathbf{B}(t) \cdot \left( \mathbf{S_1} \otimes \mathbb{I}_2  + \mathbb{I}_2 \otimes \mathbf{S_2} \right),
\end{align}
where $\xi > 0$ is the strength of the Ising coupling, $\mu$ is the common gyromagnetic ratio of the spins, $\mathbf{S_i} = (S_{ix}, S_{iy}, S_{iz})$, $i=1, 2$ is the spin operator corresponding to the $i$-th spin, with elements proportional to the Pauli matrices, and $\mathbb{I}_2$ is the $2\cross 2$ identity matrix.

A suitable orthonormal basis of this system \cite{unanyan2001preparation} is composed of the triplet states 
\begin{align}
    \label{eq:ONTriplet}
    \spinDownDown &= \spinDown_1 \otimes \spinDown_2, \nonumber \\
    \spinDownUpPlus &= \frac{1}{\sqrt{2}} \left( \spinDown_1 \otimes \spinUp_2 + \spinUp_1 \otimes \spinDown_2 \right), \\
    \spinUpUp &= \spinUp_1 \otimes \spinUp_2, \nonumber
\end{align}

and the singlet state
\begin{align}
    \label{eq:ONSinglet}
    \spinDownUpMinus &= \frac{1}{\sqrt{2}} \left( \spinDown_1 \otimes \spinUp_2 - \spinUp_1 \otimes \spinDown_2 \right),
\end{align}
where $\spinDownDown, \spinUpUp$ are the spin-down and spin-up states, respectively, while $\spinDownUpPlus, \spinDownUpMinus$ are two maximally entangled Bell states.
It can be easily verified that under Hamiltonian (\ref{Hamiltonian}) the singlet state is decoupled from the triplet states, while the probability amplitudes of the triplet states $\ket{\psi(t)} = \begin{bmatrix} \alpha_1(t), & \alpha_2(t), & \alpha_3(t) \end{bmatrix} ^T$ satisfy the following Schr\"{o}dinger equation ($\hbar = 1$)
\begin{align}
    \label{eq:ControlHamiltonian}
    i\dv{t}
    \begin{bmatrix}
    \alpha_1 \\ \alpha_2 \\ \alpha_3
    \end{bmatrix}
    =
    \begin{bmatrix}
        \xi - \beta_z & \frac{1}{\sqrt{2}} (\beta_x + i \beta_y) & \frac{1}{\sqrt{2}} (\beta_x + i \beta_y)\\
        \frac{1}{\sqrt{2}} (\beta_x - i \beta_y) & -\xi & \frac{1}{\sqrt{2}} (\beta_x + i \beta_y)\\
        0 & \frac{1}{\sqrt{2}} (\beta_x - i \beta_y) & \xi + \beta_z \\
    \end{bmatrix}
    \begin{bmatrix}
    \alpha_1 \\ \alpha_2 \\ \alpha_3
    \end{bmatrix},
\end{align}
where $\beta_n = \mu B_n$, with $n=x,y,z$. If the system starts from a state in the triplet subspace, like the case we study where the initial state is the spin-down state, then it remains in the triplet subspace and its evolution is described by the above equation.

If we choose a rotating transverse magnetic field \cite{unanyan2001preparation,stefanatos2019efficient} 
\begin{subequations}
\begin{align}
    \beta_x(t) &= \Omega(t) \cos{\omega t}, \\
    \beta_y(t) &= \Omega(t) \sin{\omega t},
\end{align}
\end{subequations}
and make the following population preserving transformation on the probability amplitudes
\begin{subequations}
\label{alpha_to_c}
    \begin{eqnarray}
    c_1 &=& \alpha_1 e^{-i(\omega+\xi) t}, \\
    c_2 &=& \alpha_2 e^{-i\xi t}, \\
    c_3 &=& \alpha_3 e^{i(\omega-\xi) t},
    \end{eqnarray}
\end{subequations}
then Eq. (\ref{eq:ControlHamiltonian}) is transformed to 
\begin{align}
    \label{eq:Schrod}
    i\dv{t}
    \begin{bmatrix}
    c_1 \\ c_2 \\ c_3
    \end{bmatrix}
    =
    \begin{bmatrix}
        \Delta(t) & \frac{\Omega(t)}{\sqrt{2}} & 0 \\
        \frac{\Omega(t)}{\sqrt{2}} & 0 & \frac{\Omega(t)}{\sqrt{2}} \\
        0 & \frac{\Omega(t)}{\sqrt{2}} & 4\xi - \Delta(t)
    \end{bmatrix}
    \begin{bmatrix}
    c_1 \\ c_2 \\ c_3
    \end{bmatrix},
\end{align}
where $\Delta(t) = 2\xi+\omega-\beta_z(t)$ is the detuning. Note that the detuning $\Delta(t)$ and the Rabi frequency $\Omega(t)$ are the two control functions used to alter the state of the system.
As can be seen from equation (\ref{eq:Schrod}), the Rabi frequency affects the transfer of population between different energy levels, while detuning changes the separation from the resonance frequency. Zero detuning in a two-level system means that it is on exact resonance, and the frequency of the pulses matches the frequency of the energy difference between the two energy levels. An example of an exact resonance pulse is the $\pi$-pulse in a qubit, which drives the system from one state to the other in minimum time \cite{d2021introduction}. But there are other quantum control methods where the detuning is varied, for example rapid adiabatic passage, where robust population inversion is achieved by keeping the Rabi frequency constant and changing the detuning of the system linearly with time, from a high negative to a high positive value \cite{shore2011manipulating}.

\subsection{Entanglement quantification through concurrence}

There are several measures which can be used to quantify the entanglement of a system. One such measure is the concurrence \cite{hill1997entanglement}, which can be expressed for pure states, like the current case, as well as for mixed states. Concurrence $C$ is bounded as follows
\begin{equation}
0\leq \mathcal{C} \leq 1,
\end{equation}
where $\mathcal{C}=0$ means that the composite quantum state is not entangled and can be written as tensor product of quantum states of the constituent subsystems, while $\mathcal{C}=1$ means that the quantum state is maximally entangled. 

We can calculate concurrence for our coupled spin pair following Ref. \cite{akhtarshenas2005concurrence}.
Any state $\ket{\psi}$ in the triplet subspace can be expressed as
\begin{align}
    \ket{\psi} = c_1 \spinDownDown + c_2 \spinDownUpPlus + c_3 \spinUpUp.
\end{align}
The state with complex conjugate probability amplitudes is
\begin{align}
    \ket{\psi^*} = c_1^* \spinDownDown + c_2^* \spinDownUpPlus + c_3^* \spinUpUp, 
\end{align}
and we use it to define
\begin{align}
    \ket{\tilde{\psi}} &= (\sigma_y \otimes \sigma_y) \ket{\psi^*} \nonumber \\
    &= - c_3^* \spinDownDown + c_2^*\spinDownUpPlus - c_1^* \spinUpUp.
\end{align}
Then, the concurrence metric is defined as:
\begin{align}
    \label{eq:concurrence}
    \mathcal{C}(\ket{\psi}) = \abs{\braket{\psi}{\tilde{\psi}}} = \abs{2c_1 c_3 - c_2^2},
\end{align}
and this is the expression that will be used in the current context to quantify entanglement in the effective three level system.

\subsection{The optimal control problem}

In previous works, adiabatic passage \cite{unanyan2001preparation} as well as shortcuts to adiabaticity \cite{paul2016high,stefanatos2019efficient} were employed to design the controls $\Omega(t), \Delta(t)$
for transferring population from the initial spin-down state $\spinDownDown$ to the maximally entangled Bell state $\ket{\psi^{\downarrow \uparrow}_+}$. Additionally, numerical optimal control was used in Ref. \cite{stefanatos2019efficient}
to find the control functions which maximize the fidelity of the Bell state at the final time $t=T$
\begin{align}
    \label{eq:fidelity1}
    \mathcal{F} = \abs{\braket{\psi(T)}{\psi^{\downarrow \uparrow}_+}}^ 2.
\end{align}
To obtain smooth control functions, which is usually easier to realize experimentally, the following truncated trigonometric series were used
\begin{eqnarray}
  \Omega(t) &=& a_0 + \sum_{k=1}^p (a_{2k-1} \cos{kt} + a_{2k} \sin{kt}), \label{TSOA_O} \\
  \Delta(t) &=& b_0 + \sum_{k=1}^p (b_{2k-1} \cos{kt} + b_{2k} \sin{kt}), \label{TSOA_D}
\end{eqnarray}
where p denotes the number of harmonics. Since in practical experiments it is 
not feasible to use unbounded controls, the control functions were bounded as follows
\begin{equation}
\label{bounds}
-\Omega_{max}\leq \Omega(t) \leq \Omega_{max}, \quad -\Delta_{max}\leq \Delta(t) \leq \Delta_{max},
\end{equation}
where the bounds $\Omega_{max}, \Delta_{max}$ are dictated by experimental limitations. We point out that the control system (\ref{eq:Schrod}) is fully controllable. First note that the Lie algebra generated by the drift matrix $M_d=\text{diag}(0, 0, 1)$, multiplying the coupling $4\xi$, and the matrices $M_{\Delta}=\text{diag}(1, 0, -1)$ and $M_{\Omega}=\lambda_1+\lambda_6$ (Gell-Mann matrices), multiplying the controls, is $\text{Lie}\{M_d, M_{\Delta}, M_{\Omega}\}=su(3)$. Since the group $SU(3)$ is compact and connected, the negative time direction for the drift term can be generated by propagating long enough in the forward direction (recurrent drift), while for the control terms by simply selecting negative values of the corresponding controls. This means that the orbits of system (\ref{eq:Schrod}) agree with the reachable sets and, along with the Lie algebra condition, we conclude that it is fully controllable.

Observe that under controls (\ref{TSOA_O}), (\ref{TSOA_D}), both the Bell state fidelity $\mathcal{F}(T)$ and the concurrence $\mathcal{C}(T)$ at the final time $t=T$ are functions of the parameter vectors $\boldsymbol{a}=\begin{pmatrix} a_0, & a_1,\ldots, & a_{2p} \end{pmatrix} ^T$ and $\boldsymbol{b}=\begin{pmatrix} b_0, & b_1,\ldots, & b_{2p} \end{pmatrix} ^T$. In the present work, we use two optimization methods, PSO and SLSQP, to find the optimal parameters which maximize the concurrence $\mathcal{C}(T)$ at the final time, under constraints (\ref{bounds}). We use a common upper bound of the controls $\Omega_{max}=\Delta_{max}$, and solve the problem for various values of this bound. For comparison, we also solve the problem of maximizing the final fidelity of the Bell state $\spinDownUpPlus$.
In addition, we also use a DL method to find the optimal controls maximizing $\mathcal{C}(T), \mathcal{F}(T)$ without assuming the ansatz (\ref{TSOA_O}), (\ref{TSOA_D}) for $\Omega(t), \Delta(t)$ but only under the constraints (\ref{bounds}).

Before moving to describe in detail the methods used in this work, we clarify the relation between the minimum times $T_F, T_C$ needed to reach $\mathcal{F}=1$ for the Bell state and $\mathcal{C}=1$ for the concurrence, respectively, under the same control constraints. Note that the target Bell state is a maximally entangled state and has the maximum possible concurrence value $\mathcal{C\left(\spinDownUpPlus\right)}=1$, as obtained from Eq. (\ref{eq:concurrence}) with $c_2=1$ and $c_1=c_3=0$, the probability amplitudes corresponding to the Bell state. But by definition $T_C$ is the minimum time to reach the level $\mathcal{C}=1$, we thus conclude that $T_C\leq T_F$. The system arrives faster at the target set of maximally entangled states $\mathcal{C} = \abs{2c_1 c_3 - c_2^2}= 1$, at a point different than the point $c_2=1, c_1=c_3=0$ corresponding to the Bell state. The methods employed find the optimal controls which drive the system to the target manifold in minimum time. For different control bounds, a different point on the manifold of maximally entangled states is reached in minimum time.

\section{Methods}
\label{sect:optim}

In this section we present the methods used for solving numerically the optimal control problem under consideration. We emphasize that both the PSO and SLSQP optimization methods use the ansatz (\ref{TSOA_O}), (\ref{TSOA_D}) for the controls and the constraints (\ref{bounds}), while the DL method utilizes only the latter constraints. 

\subsection{Particle Swarm Optimization}

PSO is a heuristic global optimization algorithm \cite{eberhart1995particle,parsopoulos2002particle} proposed by Kennedy and Eberhart. This algorithm is part of Swarm Intelligence methods, which is a subset of the evolutionary computation methods \cite{zhang2015comprehensive}. These methods study the collective behavior of decentralized, self-organized systems (natural or artificial), inspired by biological systems. Self-organization in these systems is the process where global order arises out of local interactions between the components of the initially disorganized system. This is a spontaneous process, not guided by any agent inside or outside of the system. There are only some set of rules that the components follow so that global order emerges. These rules combine strong balancing exploitation and exploration and multiple particle interactions, through which the swarm particles ``communicate" with neighbors thus information is eventually spread over the whole network. This type of rules can be formalized as an optimization algorithm which is able to solve constrained optimization problems, like the one under investigation. The constraints in the context of the PSO are integrated into the algorithm in the form of penalties. This means that particles with solutions that do not satisfy the constraints will be penalized with high costs, while those with solutions that satisfy the constraints will be rewarded with smaller costs, always respecting the objective of the problem. The pseudocode of the algorithm is presented below in Algorithm \ref{alg:PSO}.

\begin{algorithm} [H]
\caption{Particle Swarm Optimization (PSO) \cite{zhang2015comprehensive}}\label{alg:PSO}
\begin{algorithmic}
\State Initialize number of particles n and target minimum objective value
\State Initialize the inertia parameter $w$ with the $w_{max}$ and max\_iterations \\
\For{\texttt{each particle}}
    \State Initialize position X and velocity V
    \State Initialize particle best position randomly
    \State Initialize best objective value with a very high value (minimization)
\EndFor \\
\State Initialize global best position randomly \\
\State $iteration = 0$
\While{\texttt{minimum objective not reached}} 
    \State $iteration \gets iteration + 1$\\ 
    \For{\texttt{each particle}}
        \State Calculate the objective value for current position
        \State Update the personal particle best with current position if it is
        \State Update the global best with current position if needed
    \EndFor \\
    \State Calculate the new inertia parameter w:
    \State \State $w \gets w - iteration * \frac{(w_{max} - w_{min})}{max\_iterations}$ \\
    
    \For{\texttt{each particle}}
    \State Initialize arrays $r_1$ and $r_2$ randomly
        \State Calculate the new particle velocity: \\
        \State $V \gets w \cdot V + c_1 r_1 (X_{personal\_best} - X_{current}) + c_2 r_2 (X_{global\_best} - X_{current})$ \\
        \State Calculate the new particle position: $X \gets X + V$
    \EndFor
\EndWhile
\end{algorithmic}
\end{algorithm}

Evolutionary algorithms such as PSO may need more time to converge to the optimal values of the parameters, but have the great advantage that can be used in cost landscapes with several local minima besides the global minimum, since they converge to this global minimum and do not suffer from local minimum trapping such as gradient-based algorithms.

\subsection{Sequential Least Squares Programming}

Sequential Quadratic Programming (SQP) \cite{ma2024improved} is a popular method for solving constrained optimization problems. SQP is not a single algorithm, but a concept that many algorithms are based on \cite{boggs1995sequential}. The basic idea of SQP is to model the optimization problem at a given approximate solution by a quadratic programming subproblem and then to use the solution of this subproblem to construct a better approximation. The process generates a sequence of approximations that will potentially converge to an optimal solution. The method can be considered as an extension of Newton and quasi-Newton methods to the constrained optimization regime.
SLSQP is a variant of SQP where the quadratic programming subproblem is replaced by a linear constrained least squares subproblem \cite{Schittkowski82,kraft1988software}. 
This algorithm is particularly well-suited for solving nonlinear optimization problems with any combination of bounds, equality and inequality constraints.

\subsection{Deep Learning with Black Box Neural Network}

DL refers to a subset of machine learning methods which allow multi-layer neural networks to discover more abstract representations for raw input data \cite{lecun2015deep}. 
There are several DL models developed over the last two decades and used in many different applications.
One of the most common and general purpose model is that of deep feedforward neural networks. It consist of a series of hidden layers, stacked on top of each other. Each filter in the network takes data as input, and through a series of mathematical operations, transforms it into a more abstract representation. This allows the network to learn complex patterns in the data, including complex functions. During training, the network makes a prediction based on the input data. This prediction is compared to the actual answer, and the difference is used to calculate an error.  Backpropagation is the technique that allows the network to learn from these errors. It essentially works by tracing the error backward through the network, allowing each filter to adjust itself slightly to improve the overall accuracy \cite{rumelhart1986learning}. Backpropagation refers only to the process that computes the gradient, while stochastic gradient descent or another algorithm is employed afterwords to perform the learning \cite{goodfellow2016deep}. Detailed description of the training procedure can be found in several articles \cite{rumelhart1986learning}, as well as reviews and books \cite{lecun2015deep,goodfellow2016deep}.

DL methods are increasingly used in simulating complex physical systems. PINNs are trained by enforcing the physical laws in several points in the domain of interest \cite{karniadakis2021physics}. The basic idea behind PINNs is to approximate the solution of a dynamical system with a feedforward neural network. This network is trained by minimizing a loss function that values not only the predictions of the system but also enforces the governing physical laws and boundary conditions. In addition to simulating the dynamics of complex systems, DL techniques are also employed to efficiently control them. It is thus no surprise that DL methods have also been used to solve quantum control problems \cite{norambuena2024physics}. In that study, the authors propose a novel PINN architecture that finds optimal control functions in open quantum systems. It is a data-free inverse modeling DL approach, where simulation data are used to both predict and control the dynamical evolution of quantum systems, through the same neural network. There is also the possibility to use two different networks, one for the dynamics and another for the control fields, as in Ref. \cite{mowlavi2023optimal}, where the two networks are trained simultaneously with a composite loss function.

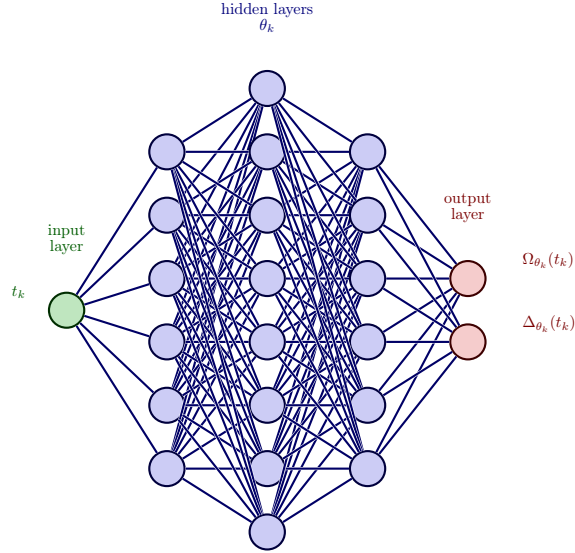
\begin{figure}[H]
    \centering
    \begin{tikzpicture}[x=2.2cm,y=1.4cm, thick, scale=0.6, every node/.style={scale=0.6}]
     \readlist\Nnod{1,6,8,6,2} 
     \message{^^J Layer}
     \foreachitem \N \in \Nnod{ 
      \def\lay{\Ncnt} 
      \pgfmathsetmacro\prev{int(\Ncnt-1)} 
      \message{\lay,}
      \foreach \i [evaluate={\y=\N/2-\i; \x=\lay; \n=\nstyle;}] in {1,...,\N}{ 
       
       \node[node \n] (N\lay-\i) at (\x,\y) {};
       
       \ifnum\lay>1 
        \foreach \j in {1,...,\Nnod[\prev]}{ 
         \draw[connect,white,line width=1.2] (N\prev-\j) -- (N\lay-\i);
         \draw[connect] (N\prev-\j) -- (N\lay-\i);
        }
       \fi 
       
      }
     }
     
     \node[above=0.3,align=center,mygreen!60!black] at (N1-1.90) {input \\[-0.2em]layer};
     \node[above=0.3,align=center,myblue!60!black] at (N3-1.90) {hidden layers\\[-0.2em]$\theta_k$};
     \node[above=0.3,align=center,myred!60!black] at (N\Nnodlen-1.90) {output\\[-0.2em]layer};
    
     \node[left=0.2,align=center,mygreen!60!black] at (N1-1.95) {$t_k$};
    
     \node[right=0.3,align=center,myred!60!black] at (N\Nnodlen-1.95) {$\Omega_{\theta_k}(t_k)$};
     \node[right=0.3,align=center,myred!60!black] at (N\Nnodlen-2.95) {$\Delta_{\theta_k}(t_k)$};
     
    \end{tikzpicture}
  \caption{BBNN. The model takes time step $t_k$ as input and outputs the values of the control functions at the input time step, $\Omega(t_k)$ and $\Delta(t_k)$.}
  \label{fig:bbnn}
\end{figure}

In the present study the system of interest is a closed quantum system whose dynamics is governed by Schr\"{o}dinger Eq. (\ref{eq:Schrod}), thus there is no need to simulate the underlying physics. For this reason, here we use a simplified version of the PINN model introduced in Ref. \cite{norambuena2024physics}, which does not predict the state evolution but only enforces the objective function at the final time step. The model is a Black Box Neural Network (BBNN) shown in Fig. \ref{fig:bbnn}, that takes as input each discretized time in the evolution interval and outputs the optimal values of the control functions at the input time step. 
The model essentially implements a constrained optimization since the predicted control function values are restricted between the bounds (\ref{bounds}).

The objective functions to be minimized, used for the BBNN training, depend only on the quantum metric used to evaluate the final quantum state, which are the concurrence $\mathcal{C}$ (\ref{eq:concurrence}) or the Bell state fidelity $\mathcal{F}$ (\ref{eq:fidelity1}), 
\begin{align}
    J(\ket{\psi(T)}, \theta) = 1 - \mathcal{C}(\ket{\psi(T)}, \theta)
\end{align}
and
\begin{align}
    J(\ket{\psi(T)}, \theta) = 1 - \mathcal{F} (\ket{\psi(T)}, \theta),
\end{align}
respectively, while $\theta$ are the model parameters.

Initially, the time is discretized in time steps with no uncertainty assumed and each time step is fed into the BBNN which outputs some random initial control function values, since the initial parameters $\theta$ of the model are randomly selected. In each iteration, the model produces the control values in each time step which are used to evolve the quantum system in time. Subsequently, the resultant final state is evaluated with the cost function used, and based on the result the parameters of the model are updated by the backpropagation and optimization algorithm (Adam optimizer). After a certain number of epochs, the model converges to a parameter set producing control functions which drive the quantum system to a final state maximizing the quantum metric used in the training.

\section{Results and discussion}
\label{sect:results}

We use the presented optimization methods PSO and SLSQP to find the optimal parameters in the trigonometric expansions (\ref{TSOA_O}), (\ref{TSOA_D}) of the controls with $p=5$ harmonics, under constraint (\ref{bounds}) and for different control bounds. We also use the presented BBNN to find the optimal controls $\Omega(t), \Delta(t)$, without assuming the forms (\ref{TSOA_O}), (\ref{TSOA_D}), but still under constraint (\ref{bounds}).


The parameters of the PSO algorithm are shown in Table \ref{Tab:PSOParams}. The algorithm has been implemented using Python programming language and the mean execution time is about $15$ minutes. The code will be available in a GitHub repository publicly available.

\begin{table}[ht]
    \begin{tabular}{ | m{5cm} <{\centering} | m{2cm} <{\centering} |} 
      \hline
      \textbf{Parameter} & Value \\
      \hline
      Nb particles & 300 \\
      \hline
      Max iterations & 600 \\
      \hline
      max inertia ($w_{max}$) & 0.9 \\
      \hline
      min inertia ($w_{max}$) & 0.2 \\
      \hline
      $c_1$ & 1 \\
      \hline
      $c_2$ & 3 \\
      \hline
    \end{tabular}
    \caption{PSO algorithm parameters}
    \label{Tab:PSOParams}
\end{table}

\begin{figure}[H]
\centering
    \includegraphics[scale=0.4]{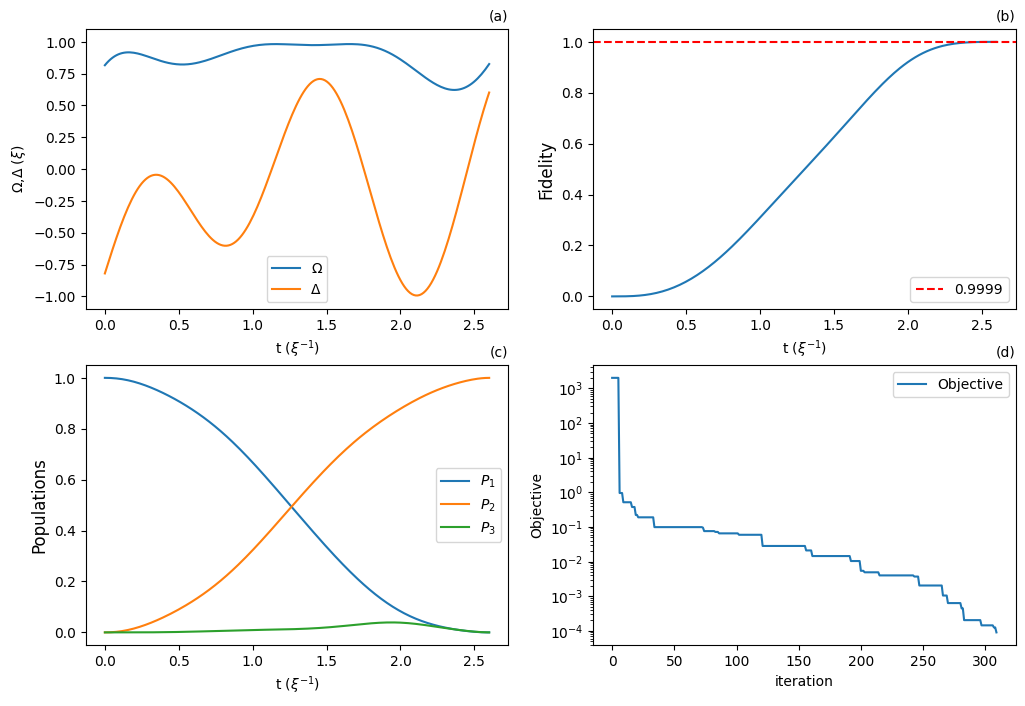}
    \caption{Results for PSO stochastic optimization achieving fidelity $\mathcal{F}>0.9999$ for state transfer $\spinDownDown \rightarrow \spinDownUpPlus$ with control bound $\Omega_{max}/\xi=\Delta_{max}/\xi=1$: (a) Optimal Rabi frequency $\Omega(t)$ and detuning $\Delta(t)$, (b) Fidelity (population of state $\spinDownUpPlus$), (c) Populations of states $\spinDownDown$, $\spinDownUpPlus$ and $\spinUpUp$, (d) Objective function $1-\mathcal{C}$.}
    \label{fig:PSOBound1}
\end{figure}

\begin{figure}[H]
\centering
    \includegraphics[scale=0.4]{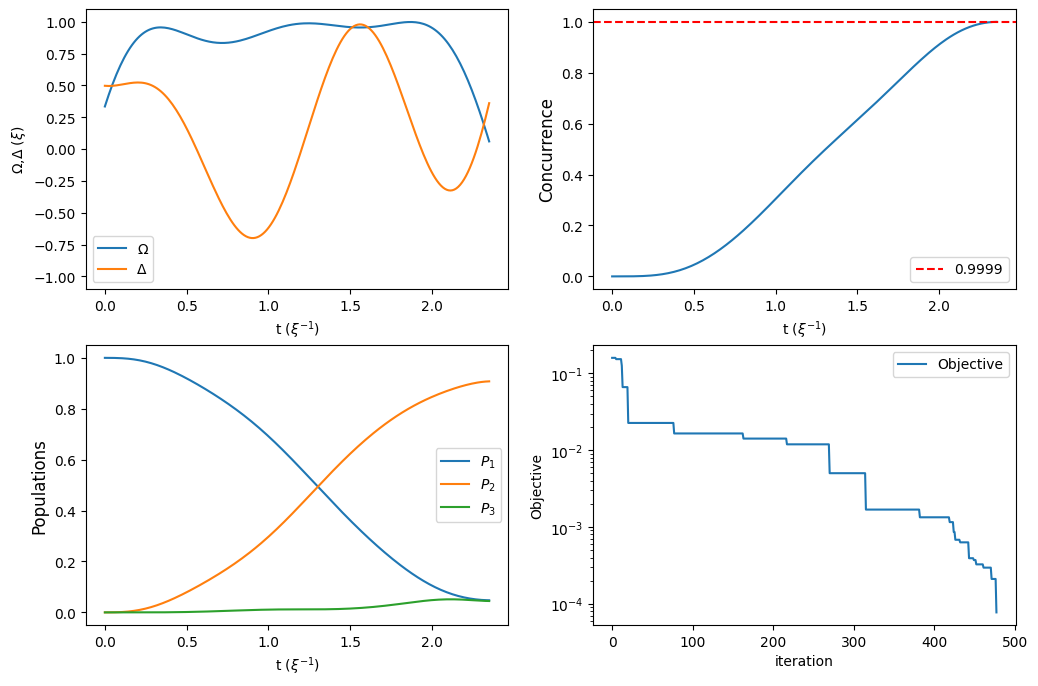}
    \caption{Results for PSO stochastic optimization achieving concurrence $\mathcal{C}>0.9999$ with control bound $\Omega_{max}/\xi=\Delta_{max}/\xi=1$: (a) Optimal Rabi frequency $\Omega(t)$ and detuning $\Delta(t)$, (b) Concurrence, (c) Populations of states $\spinDownDown$, $\spinDownUpPlus$ and $\spinUpUp$, (d) Objective function $1-\mathcal{C}$.}
    \label{fig:PSOMaxConcDetailed}
\end{figure}

\begin{figure}[H]
\centering
    \includegraphics[scale=0.4]{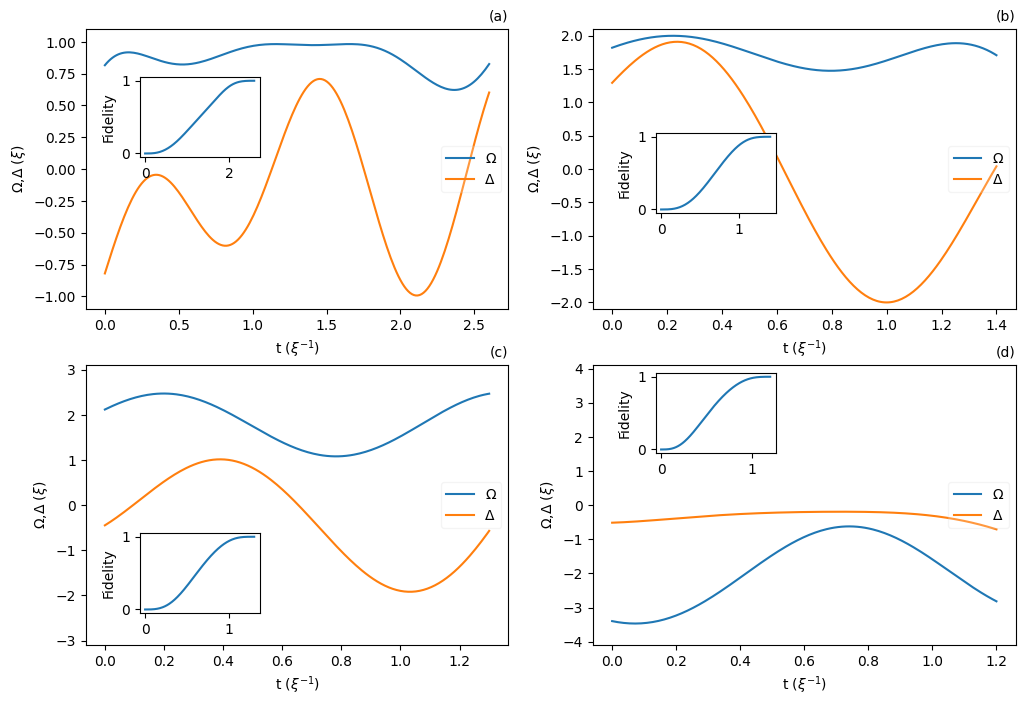}
    \caption{Optimal controls $\Omega(t), \Delta(t)$ obtained with PSO stochastic optimization and achieving fidelity $\mathcal{F}>0.9999$ for state transfer $\spinDownDown \rightarrow \spinDownUpPlus$, for four different values of the upper control bound: (a) $\Omega_{max}/\xi=\Delta_{max}/\xi=1$ , (b) $\Omega_{max}/\xi=\Delta_{max}/\xi=2$, (c) $\Omega_{max}/\xi=\Delta_{max}/\xi=3$, (d) $\Omega_{max}/\xi=\Delta_{max}/\xi=4$. The inset in each subfigure highlights the corresponding time evolution of fidelity.}
    \label{fig:PSOFidelityComp}
\end{figure}

\begin{figure}[H]
\centering
    \includegraphics[scale=0.4]{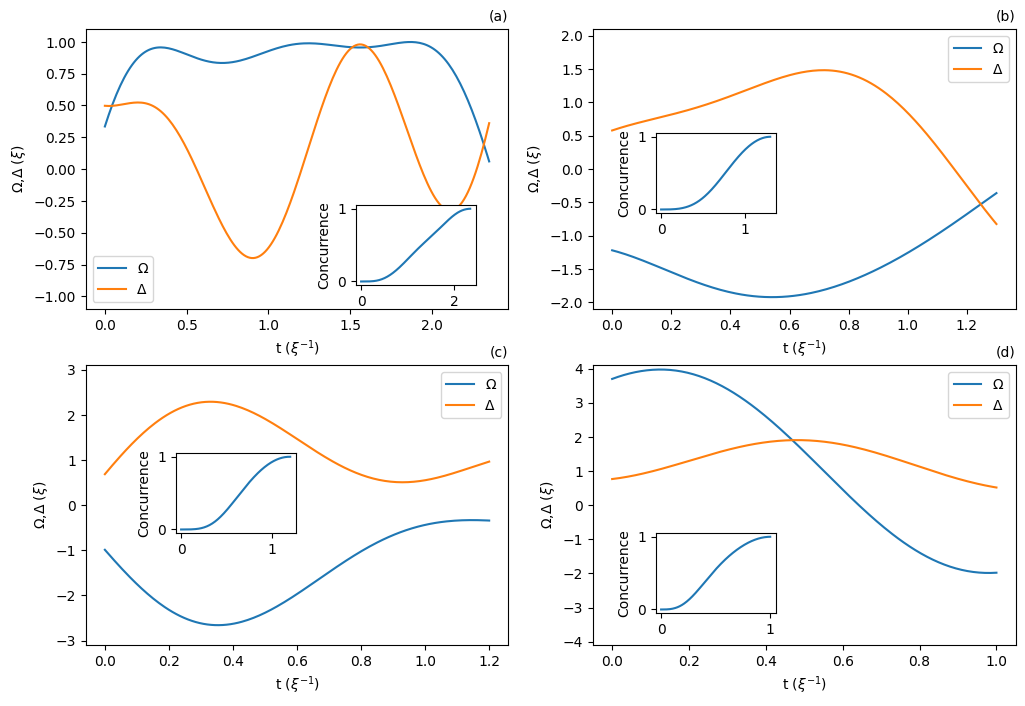}
    \caption{Optimal controls $\Omega(t), \Delta(t)$ obtained with PSO stochastic optimization and achieving concurrence $\mathcal{C}>0.9999$, for four different values of the upper control bound: (a) $\Omega_{max}/\xi=\Delta_{max}/\xi=1$ , (b) $\Omega_{max}/\xi=\Delta_{max}/\xi=2$, (c) $\Omega_{max}/\xi=\Delta_{max}/\xi=3$, (d) $\Omega_{max}/\xi=\Delta_{max}/\xi=4$. The inset in each subfigure highlights the corresponding time evolution of concurrence.}
    \label{fig:PSOMaxConcComp}
\end{figure}

\begin{figure}[H]
\centering
    \includegraphics[scale=0.6]{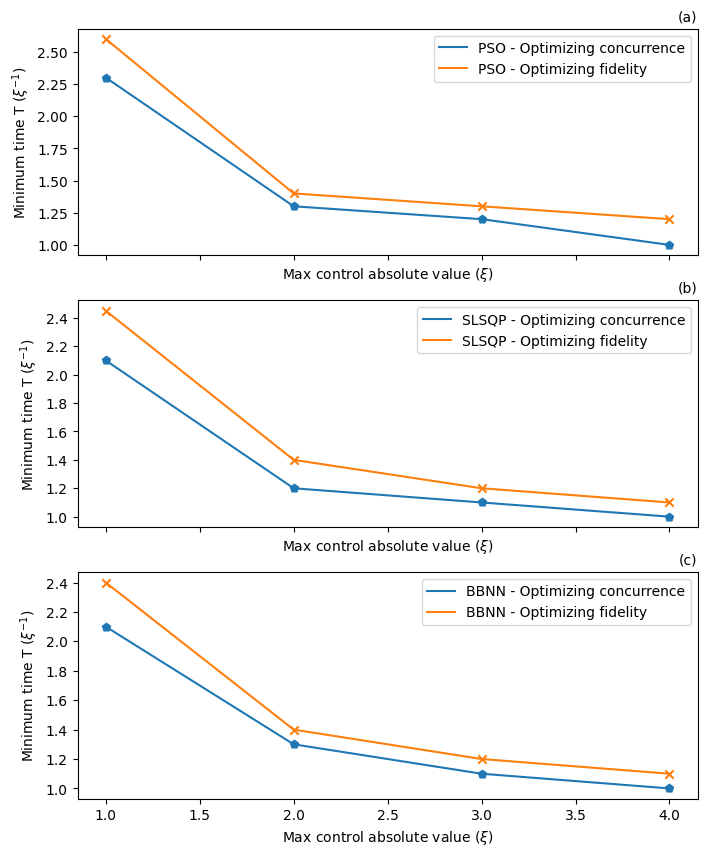}
    \caption{Minimum time T to reach the target level ($0.9999$) of fidelity (yellow line) and concurrence (blue line), as a function of the maximum absolute control value, with each method used: (a) PSO, (b) SLSQP, (c) BBNN.}
    \label{fig:FinalTimingsPSO}
\end{figure}

\begin{figure}[H]
\centering
    \includegraphics[scale=0.6]{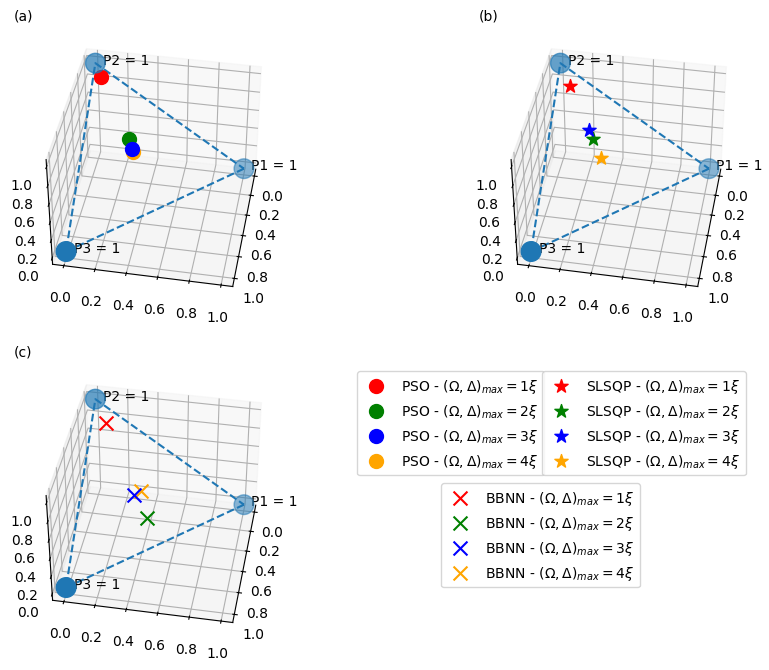}
    \caption{Final states with concurrence $\mathcal{C}>0.9999$, represented on the two-dimensional simplex defined by the populations of the constituent triplet basis states, for the three methods used and four maximum control absolute values $\Omega_{max}/\xi=\Delta_{max}/\xi=1 (\mbox{red}), 2 (\mbox{green}), 3 (\mbox{blue}), 4 (\mbox{yellow})$. A different subfigure and marker is used for each method: (a) PSO (filled circles), (b) SLSQP (stars), (c) BBNN (crosses).}
    \label{fig:Simplex}
\end{figure}

To fix ideas and also create a basis for comparison, we start with maximizing the fidelity $\mathcal{F}$ of the Bell state $\ket{\psi^{\downarrow \uparrow}_+}$, as in Ref. \cite{stefanatos2019efficient}, for control bounds $\Omega_{max}/\xi=\Delta_{max}/\xi=1$. The results are displayed in Fig. \ref{fig:PSOBound1}, where the terminal time T used for the quantum evolution is the minimum time achieving the desired fidelity level $\mathcal{F} > 0.9999$. In Fig. \ref{fig:PSOBound1}(a) we show the optimal controls $\Omega(t)$ (blue line) and $\Delta(t)$ (yellow line), in Figs. \ref{fig:PSOBound1}(b,c) the time evolution of the fidelity and the populations of the triplet states (\ref{eq:ONTriplet}), while in Fig. \ref{fig:PSOBound1}(d) the change of the objective function $1-\mathcal{F}$ with the number of iterations. The optimal coefficients of the truncated trigonometric series (\ref{TSOA_O}), (\ref{TSOA_D}) used for the controls are given in Table \ref{Tab:TSOAParamsFidTablePSO}. Similar results are displayed for concurrence in Fig. \ref{fig:PSOMaxConcDetailed} and Table \ref{Tab:TSOAParamsConcTable}, where levels $\mathcal{C}>0.9999$ are obtained using the same control bounds $\Omega_{max}/\xi=\Delta_{max}/\xi=1$. Observe that the target concurrence level is obtained in less than $2.5$ time units, faster than the same Bell state fidelity level which is achieved in more than $2.5$ time units. Also, the final state is different from the Bell state $\spinDownUpPlus$, see the final population $P_2$ in Fig. \ref{fig:PSOMaxConcDetailed}(c) which is appreciably smaller than unity. In Fig. \ref{fig:PSOFidelityComp} we present results regarding Bell state fidelity maximization for four different upper control bounds $\Omega_{max}/\xi=\Delta_{max}/\xi=1, 2, 3, 4$. For each bound value we display the corresponding optimal controls and the fidelity evolution in the inset. The terminal time T of the quantum evolution in each case is the minimum time needed to obtain the desired fidelity level $\mathcal{F} > 0.9999$. Observe that, as the upper control bound increases, the minimum time to reach the Bell state decreases and the shape of the control functions changes. In Fig. \ref{fig:PSOMaxConcComp} we show analogous results for maximizing the concurrence. Comparing with Fig. \ref{fig:PSOFidelityComp}, we observe that in all cases the target concurrence level $\mathcal{C}>0.9999$ is obtained faster than the same level of the Bell state fidelity. This is better illustrated in Fig. \ref{fig:FinalTimingsPSO}(a), where we plot the necessary minimum time for each case as a function of the maximum control absolute value. Another interesting observation is that, as the control bound changes, the maximally entangled state reached when maximizing concurrence also changes. This is demonstrated in Fig. \ref{fig:Simplex}(a), where we place the final state reached for each control bound on the two-dimensional simplex defined by the constituent populations of the triplet states, $\sum_{i=1}^3P_i=1$ and $P_i\geq 0$, while note that the specific Bell state corresponds to the upper corner of the triangle. 


\begin{table}[ht]
    \begin{tabular}{ | m{1cm} <{\centering} | m{4cm} <{\centering} | m{4cm} <{\centering} |} 
      \hline
      \textbf{i} & \boldmath$\Omega: a_i$ & \boldmath$\Delta: b_i$ \\
      \hline
      0 & 0.16472369 & - 0.46195841 \\
      \hline
      1 & -0.19892197 & - 0.64841354 \\
      \hline
      2 & 1 & 0.66455828 \\
      \hline
      3 & -0.0687911 & 0.47527272 \\
      \hline
      4 & 0.21946375 & 0.45552122 \\
      \hline
      5 & 0.09341737 & - 0.10586344 \\
      \hline
      6 & 0.89075888 & -0.4108748 \\
      \hline
      7 & 1 & -0.29944072 \\
      \hline
      8 & -0.21510914 & -0.0954528 \\
      \hline
      9 & -0.17416499 & 0.22079337 \\
      \hline
      10 & -0.36284854 & 0.7472183 \\
      \hline
    \end{tabular}
    \caption{Optimal parameters obtained with the PSO algorithm for the truncated trigonometric series (\ref{TSOA_O}), (\ref{TSOA_D}) used for the controls, achieving fidelity $\mathcal{F}>0.9999$ for the Bell state $\spinDownUpPlus$ with control bound $\Omega_{max}/\xi=\Delta_{max}/\xi=1$.}
    \label{Tab:TSOAParamsFidTablePSO}
\end{table}

\begin{table}[ht]
    \begin{tabular}{ | m{1cm} <{\centering} | m{4cm} <{\centering} | m{4cm} <{\centering} |} 
      \hline
      \textbf{i} & \boldmath$\Omega: a_i$ & \boldmath$\Delta: b_i$ \\
      \hline
      \hline
      0 & 0.12202841 &  0.03785183 \\
      \hline
      1 & -0.10813239 & -0.33607908 \\
      \hline
      2 & 0.57533857 & 0.76542096 \\
      \hline
      3 & -0.2292401 & 0.96728942 \\
      \hline
      4 & 0.69918352 & -0.05339743 \\
      \hline
      5 & -0.14302123 & 0.59587648 \\
      \hline
      6 & -0.10960869 & -0.59751802 \\
      \hline
      7 & 0.24047154 & -0.45444928 \\
      \hline
      8 & 1 & -1 \\
      \hline
      9 & 0.4529411 & -0.31287144 \\
      \hline
      10 & -0.32113074 & 1 \\
      \hline
    \end{tabular}
    \caption{Optimal parameters obtained with the PSO algorithm for the truncated trigonometric series (\ref{TSOA_O}), (\ref{TSOA_D}) used for the controls, achieving concurrence $\mathcal{C}>0.9999$ with control bound $\Omega_{max}/\xi=\Delta_{max}/\xi=1$.}
    \label{Tab:TSOAParamsConcTable}
\end{table}

\begin{figure}[H]
\centering
    \includegraphics[scale=0.4]{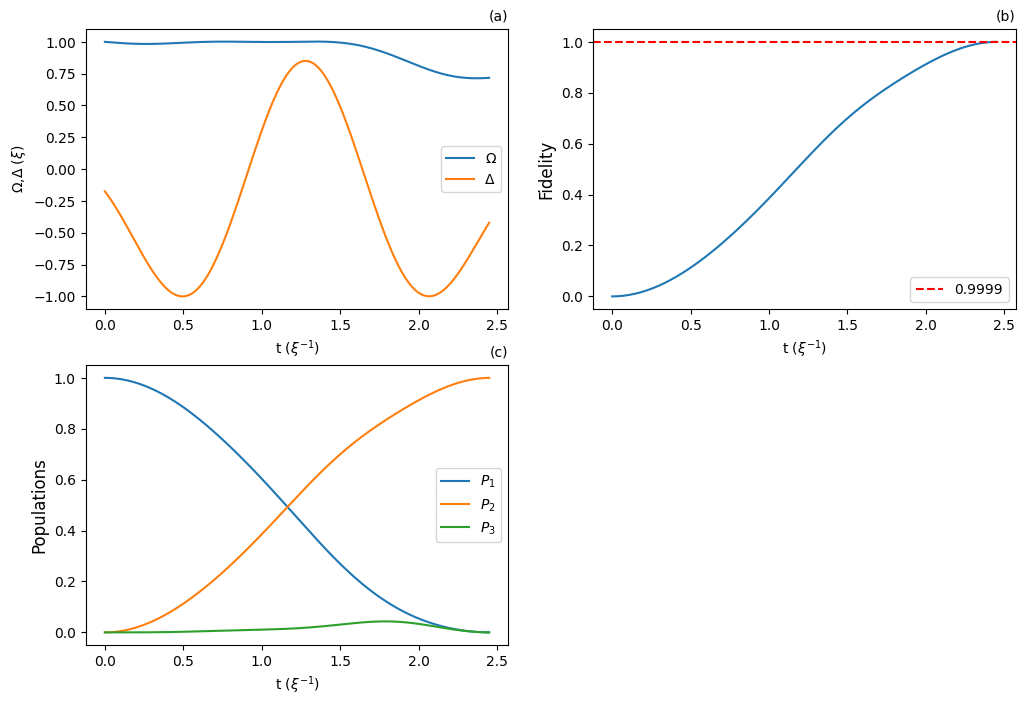}
    \caption{Results for SLSQP optimization algorithm achieving fidelity $\mathcal{F}>0.9999$ with control bound $\Omega_{max}/\xi=\Delta_{max}/\xi=1$: (a) Optimal Rabi frequency $\Omega(t)$ and detuning $\Delta(t)$, (b) Fidelity, (c) Populations of states $\spinDownDown$, $\spinDownUpPlus$ and $\spinUpUp$.}
    \label{fig:SLSQPDetailed}
\end{figure}

\begin{figure}[H]
\centering
    \includegraphics[scale=0.4]{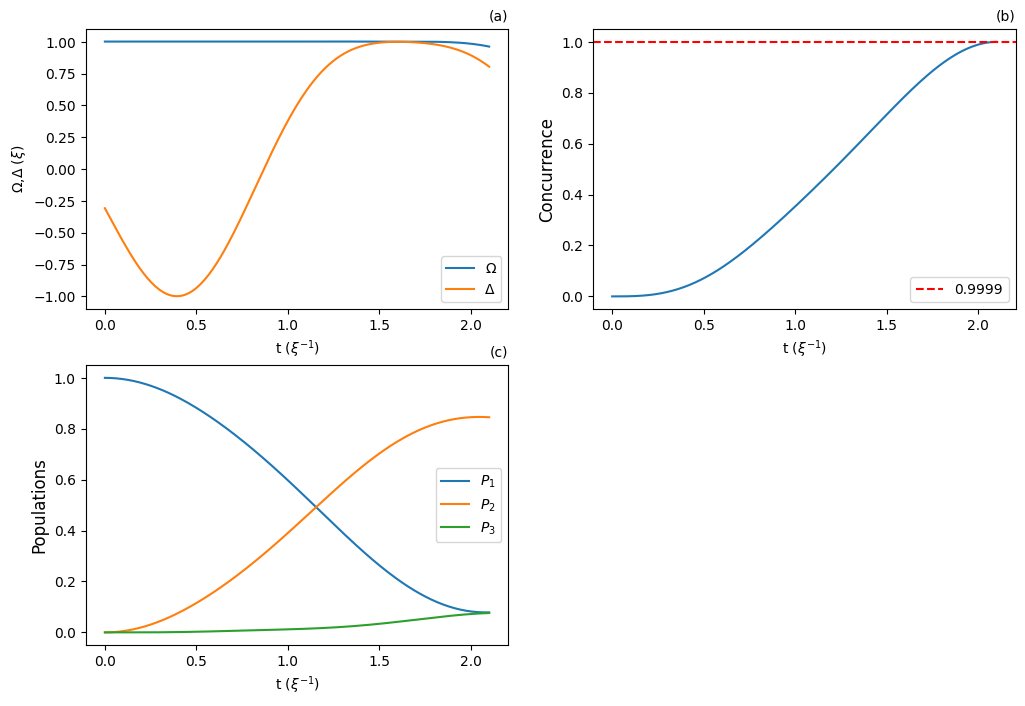}
    \caption{Results for SLSQP optimization algorithm achieving concurrence $\mathcal{C}>0.9999$ with control bound $\Omega_{max}/\xi=\Delta_{max}/\xi=1$: (a) Optimal Rabi frequency $\Omega(t)$ and detuning $\Delta(t)$, (b) Concurrence, (c) Populations of states $\spinDownDown$, $\spinDownUpPlus$ and $\spinUpUp$.}
    \label{fig:SLSQPConcDetailed}
\end{figure}

\begin{figure}[H]
\centering
    \includegraphics[scale=0.4]{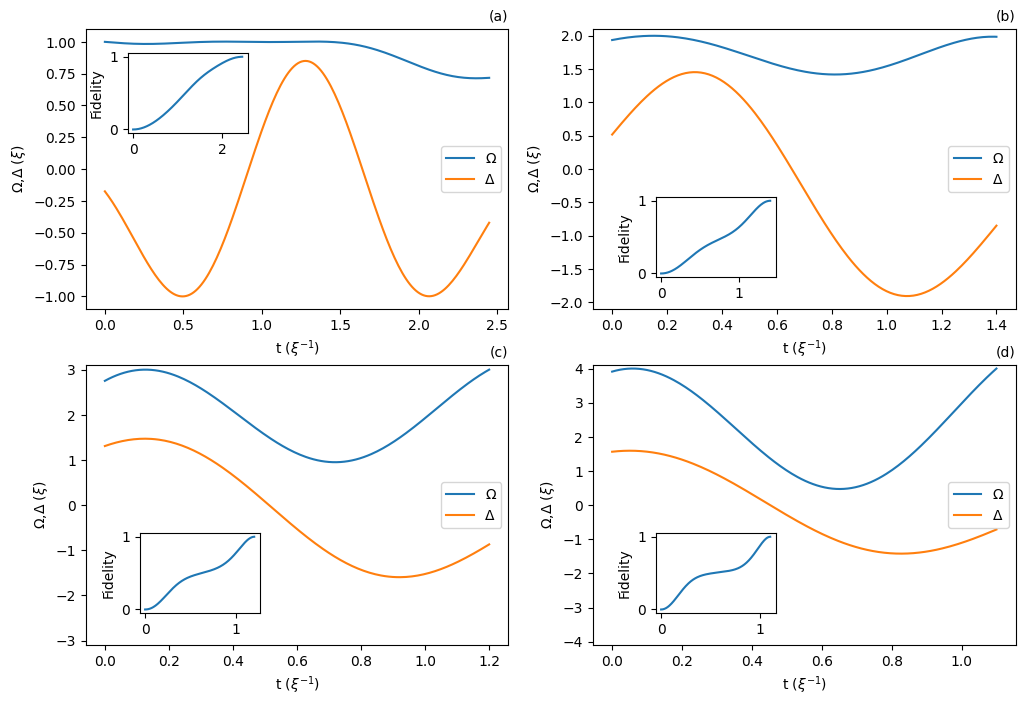}
    \caption{Optimal controls $\Omega(t), \Delta(t)$ obtained with SLSQP algorithm and achieving fidelity $\mathcal{F}>0.9999$ for state transfer $\spinDownDown \rightarrow \spinDownUpPlus$, for four different values of the upper control bound: (a) $\Omega_{max}/\xi=\Delta_{max}/\xi=1$ , (b) $\Omega_{max}/\xi=\Delta_{max}/\xi=2$, (c) $\Omega_{max}/\xi=\Delta_{max}/\xi=3$, (d) $\Omega_{max}/\xi=\Delta_{max}/\xi=4$. The inset in each subfigure highlights the corresponding time evolution of fidelity.}
    \label{fig:SLSQPFidComp}
\end{figure}

\begin{figure}[H]
\centering
    \includegraphics[scale=0.4]{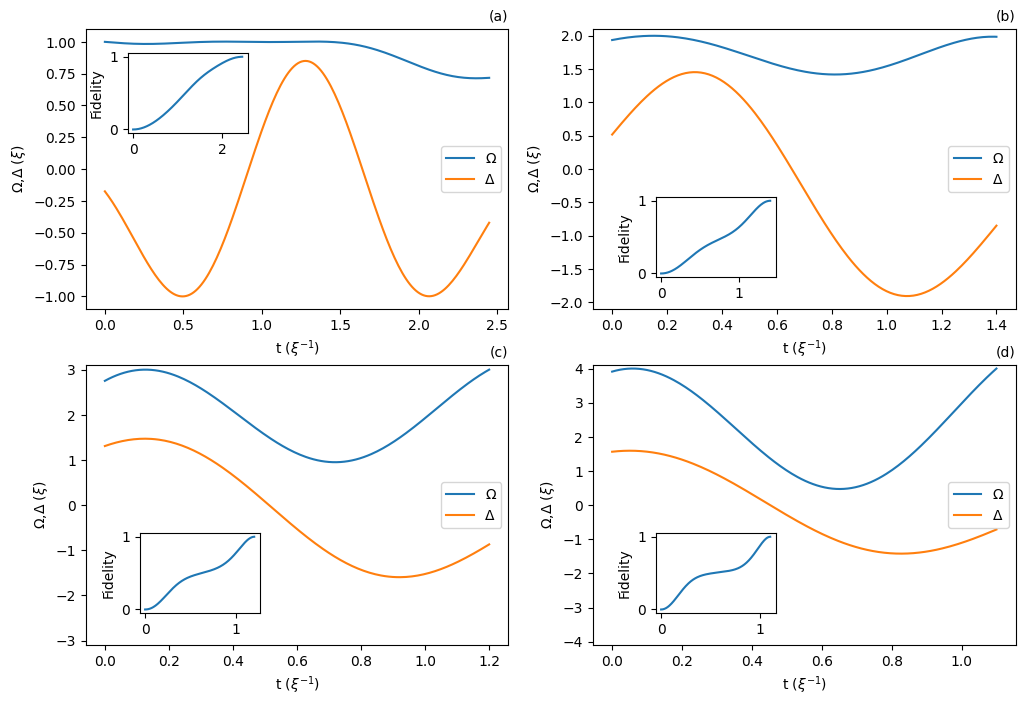}
    \caption{Optimal controls $\Omega(t), \Delta(t)$ obtained with SLSQP algorithm and achieving concurrence $\mathcal{C}>0.9999$, for four different values of the upper control bound: (a) $\Omega_{max}/\xi=\Delta_{max}/\xi=1$ , (b) $\Omega_{max}/\xi=\Delta_{max}/\xi=2$, (c) $\Omega_{max}/\xi=\Delta_{max}/\xi=3$, (d) $\Omega_{max}/\xi=\Delta_{max}/\xi=4$. The inset in each subfigure highlights the corresponding time evolution of concurrence.}
    \label{fig:SLSQPConcComp}
\end{figure}

The SLSQP algorithm has been already implemented in SciPy python scientific package \cite{2020SciPy-NMeth}. Our algorithm solving the quantum control problem under study employs the SciPy package to perform the optimization of the trigonometric series parameters so that fidelity or concurrence is maximized. The execution time of this algorithm is about 5 minutes and we obtain results similar to those with the PSO algorithm. Specifically, in Figs. \ref{fig:SLSQPDetailed} and \ref{fig:SLSQPConcDetailed} we present the optimal controls and the time evolution of populations and fidelity/concurrence, respectively, for control bound $\Omega_{max}/\xi=\Delta_{max}/\xi=1$. The corresponding optimal parameters are given in Tables \ref{Tab:TSOAParamsFidTableSLSQP} and \ref{Tab:TSOAParamsConTableSLSQP}. Observe that the desired concurrence level $\mathcal{C}>0.9999$ is again achieved faster than the same Bell state fidelity level $\mathcal{F}>0.9999$, while the maximally entangled state reached is different than the particular Bell state, since the $P_2$ population in Fig. \ref{fig:SLSQPConcDetailed}(c) is noticeably less than unity. In Figs. \ref{fig:SLSQPFidComp} and \ref{fig:SLSQPConcComp} we show the optimal controls maximizing fidelity/concurrence (also displayed in the inset), respectively, for four different values of the control bound $\Omega_{max}/\xi=\Delta_{max}/\xi=1, 2, 3, 4$. As before, in all cases the desired concurrence level is obtained faster than the targeted Bell state fidelity level, as demonstrated in Fig. \ref{fig:FinalTimingsPSO}(b), while the maximally entangled states reached for each control bound are different, as shown in Fig. \ref{fig:Simplex}(b).

\begin{table}[ht]
    \begin{tabular}{ | m{1cm} <{\centering} | m{4cm} <{\centering} | m{4cm} <{\centering} |} 
      \hline
      \textbf{i} & \boldmath$\Omega: a_i$ & \boldmath$\Delta: b_i$ \\
      \hline
      \hline
      0 & 0.62186844 & -0.19164546 \\
      \hline
      1 & 0.25751019 & 0.00622775 \\
      \hline
      2 & 0.37180072 & -0.07791177 \\
      \hline
      3 & 0.03320865 & -0.16113252 \\
      \hline
      4 & -0.00203274 & 0.08881795 \\
      \hline
      5 & 0.13143564 & -0.24381851 \\
      \hline
      6 & 0.04216778 & -0.20513285 \\
      \hline
      7 & 0.03639866 & 0.13803518 \\
      \hline
      8 & -0.18368735 & -0.30414457 \\
      \hline
      9 & -0.08042158 & 0.27782271 \\
      \hline
      10 & 0.02880343 & 0.02403051 \\
      \hline
    \end{tabular}
    \caption{Optimal parameters obtained with the SLSQP algorithm for the truncated trigonometric series (\ref{TSOA_O}), (\ref{TSOA_D}) used for the controls, achieving fidelity $\mathcal{F}>0.9999$ for the Bell state $\spinDownUpPlus$ with control bound $\Omega_{max}/\xi=\Delta_{max}/\xi=1$.}
    \label{Tab:TSOAParamsFidTableSLSQP}
\end{table}

\begin{table}[ht]
    \begin{tabular}{ | m{1cm} <{\centering} | m{4cm} <{\centering} | m{4cm} <{\centering} |} 
      \hline
      \textbf{i} & \boldmath$\Omega: a_i$ & \boldmath$\Delta: b_i$ \\
      \hline
      \hline
      0 & 0.60821182 & 0.11778919 \\
      \hline
      1 & 0.4259045 & -0.15750619 \\
      \hline
      2 & 0.32534244 & 0.23127512 \\
      \hline
      3 & -0.1210874 & -0.40484222 \\
      \hline
      4 & -0.06469339 & -0.01516328 \\
      \hline
      5 & 0.19312832 & -0.35004653 \\
      \hline
      6 & -0.03348454 & 0.11859576 \\
      \hline
      7 & -0.11433822 & -0.07917745 \\
      \hline
      8 & -0.05770413 & -0.07917745 \\
      \hline
      9 & 0.01036488 & 0.03397065 \\
      \hline
      10 & 0.02813306 & -0.22273092 \\
      \hline
    \end{tabular}
    \caption{Optimal parameters obtained with the SLSQP algorithm for the truncated trigonometric series (\ref{TSOA_O}), (\ref{TSOA_D}) used for the controls, achieving concurrence $\mathcal{C}>0.9999$ with control bound $\Omega_{max}/\xi=\Delta_{max}/\xi=1$.}
    \label{Tab:TSOAParamsConTableSLSQP}
\end{table}

The BBNN method solves a different optimization problem than the previous two methods. Specifically, the time interval is discretized in $1000$ steps and the neural network predicts directly the optimal control values for each individual time step. The time discretization is very small and the space can be characterized as pseudo-continuous due to the large number of timesteps, thus  it is not expected  to affect the results. The model is able to grasp quite fast with no need to see any previous experiments or be guided by previous examples, but learns to produce the control function values directly by simulation data. The controls are not forced to be smooth, as in the previous two cases, not even continuous. The BBNN model has been created with the help of the PyTorch package \cite{NEURIPS2019_9015}, while the quantum evolution simulations are also implemented with the PyTorch library, using rank-{1} and rank-{2} tensors to represent vectors and operators (matrices) in the associated Hilbert space. The parameters for the deep neural network used by the BBNN method are given in Table \ref{Tab:AmplConcBBNN}.
The mean execution time is about $5$ minutes.

\begin{table}[ht]
    \begin{tabular}{ | c | c | c |} 
      \hline
      Parameter & Max Fidelity & Max Concurrence \\
      \hline
      Number of hidden layers & 4 & 4 \\ 
      \hline
      Number of parameters per hidden layer & 75 & 75 \\
      \hline
      Activation function & Tanh & Sin \\
      \hline
      Optimizer & Adam & Adam \\
      \hline
      Learning rate & 1e-3 & 1e-2 \\
      \hline
    \end{tabular}
    \caption{Parameters of the deep neural network for BBNN method}
    \label{Tab:AmplConcBBNN}
\end{table}

\begin{figure}[H]
\centering
    \includegraphics[scale=0.4]{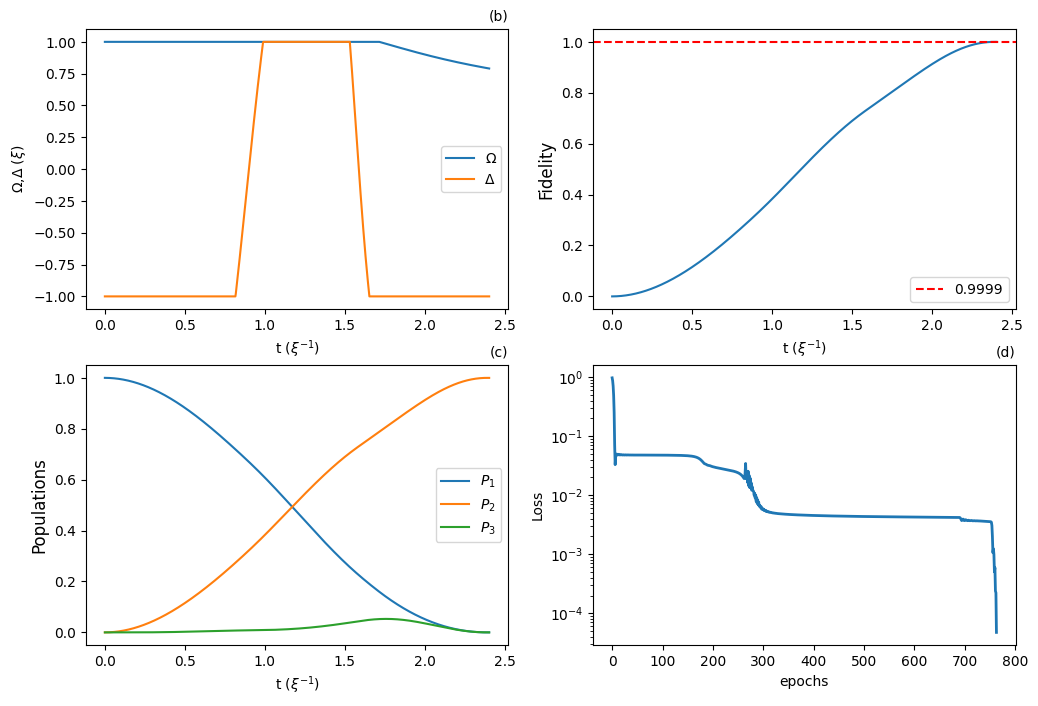}
    \caption{Results for BBNN method achieving fidelity $\mathcal{F}>0.9999$ with control bound $\Omega_{max}/\xi=\Delta_{max}/\xi=1$: (a) Optimal Rabi frequency $\Omega(t)$ and detuning $\Delta(t)$, (b) Fidelity, (c) Populations of states $\spinDownDown$, $\spinDownUpPlus$ and $\spinUpUp$, (d) Objective function $1-\mathcal{F}$.}
    \label{fig:BBNNFidDetailed}
\end{figure}

\begin{figure}[H]
\centering
    \includegraphics[scale=0.4]{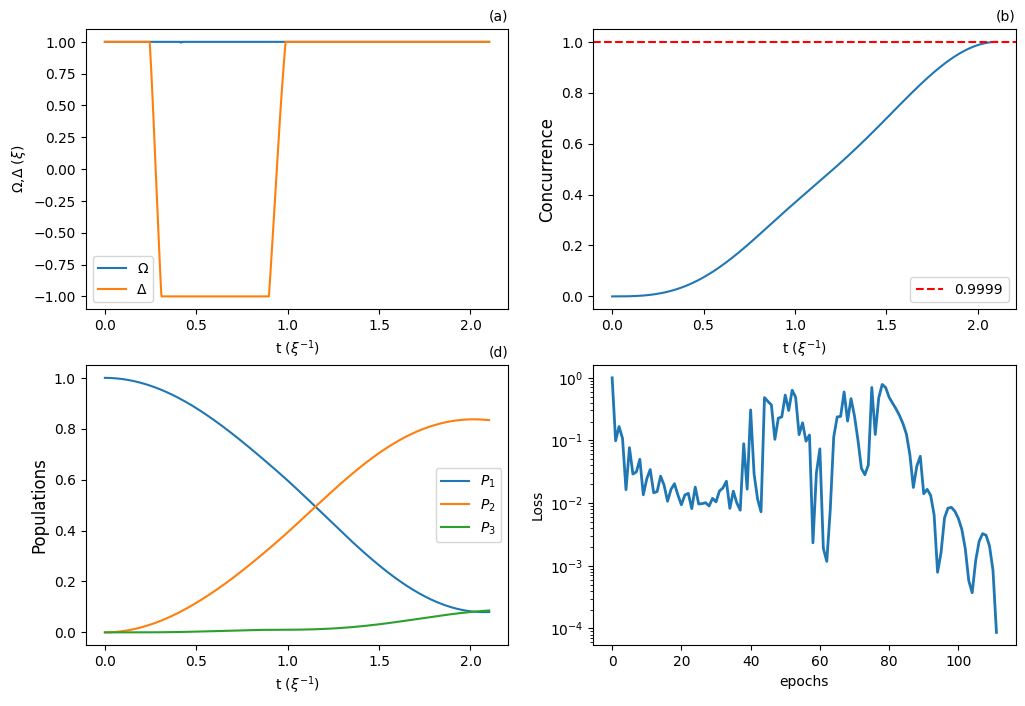}
    \caption{Results for BBNN method achieving concurrence $\mathcal{C}>0.9999$ with control bound $\Omega_{max}/\xi=\Delta_{max}/\xi=1$: (a) Optimal Rabi frequency $\Omega(t)$ and detuning $\Delta(t)$, (b) Concurrence, (c) Populations of states $\spinDownDown$, $\spinDownUpPlus$ and $\spinUpUp$, (d) Objective function $1-\mathcal{C}$.}
    \label{fig:Tab:BBNNConDetailed}
\end{figure}

\begin{figure}[H]
\centering
    \includegraphics[scale=0.4]{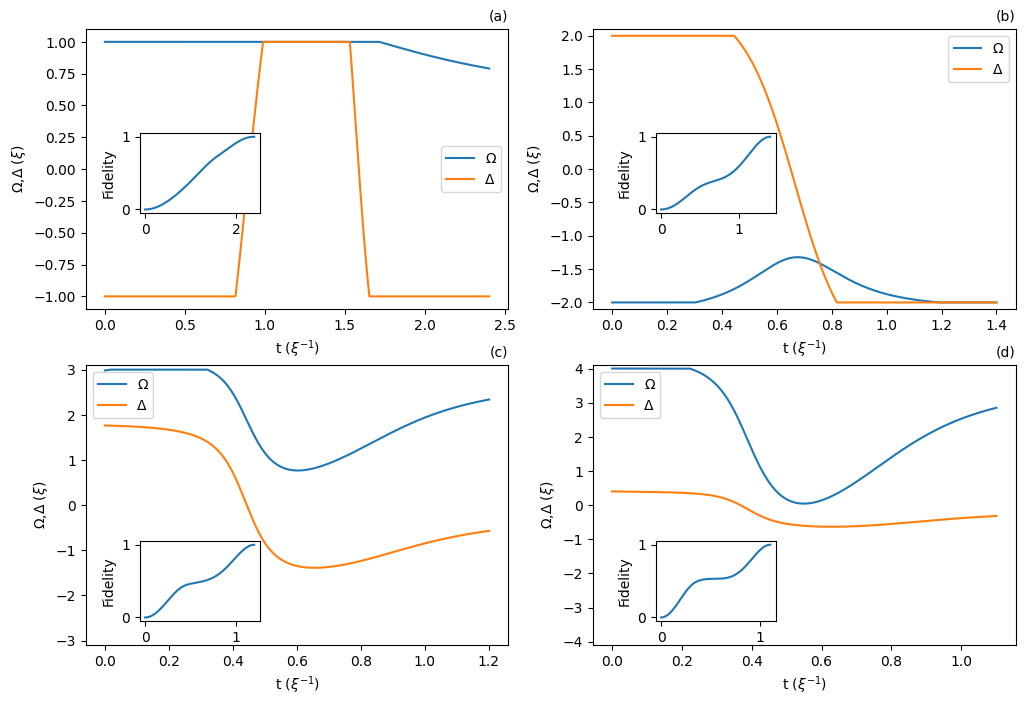}
    \caption{Optimal controls $\Omega(t), \Delta(t)$ obtained with BBNN method and achieving fidelity $\mathcal{F}>0.9999$, for four different values of the upper control bound: (a) $\Omega_{max}/\xi=\Delta_{max}/\xi=1$ , (b) $\Omega_{max}/\xi=\Delta_{max}/\xi=2$, (c) $\Omega_{max}/\xi=\Delta_{max}/\xi=3$, (d) $\Omega_{max}/\xi=\Delta_{max}/\xi=4$. The inset in each subfigure highlights the corresponding time evolution of fidelity.}
    \label{fig:BBNNFidComp}
\end{figure}

\begin{figure}[H]
\centering
    \includegraphics[scale=0.4]{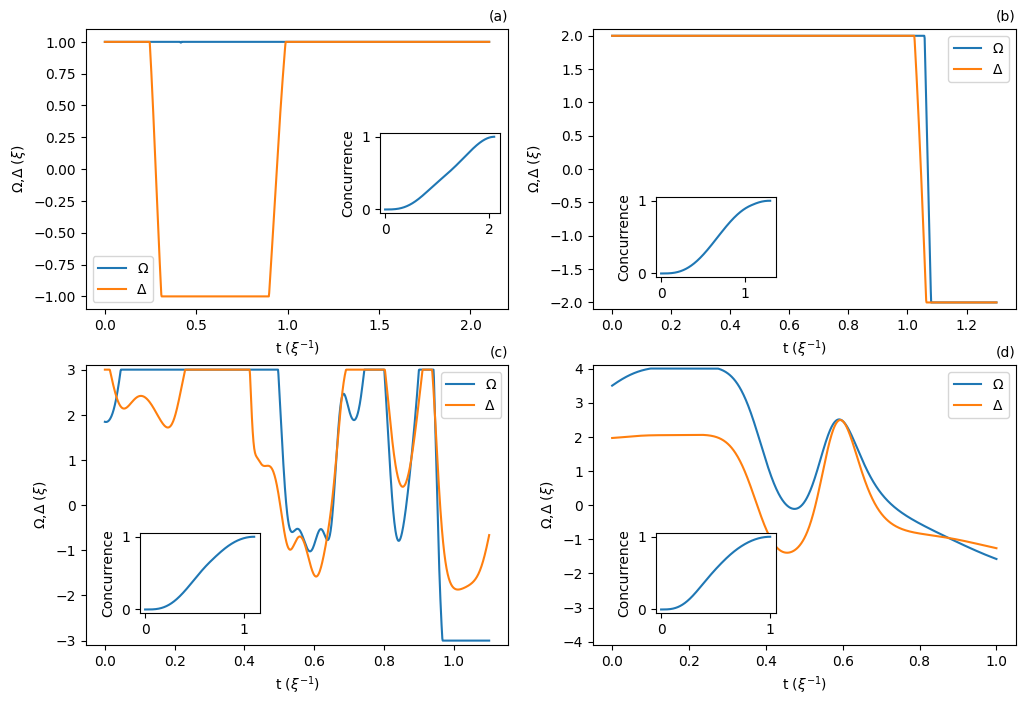}
    \caption{Optimal controls $\Omega(t), \Delta(t)$ obtained with BBNN method and achieving concurrence $\mathcal{C}>0.9999$, for four different values of the upper control bound: (a) $\Omega_{max}/\xi=\Delta_{max}/\xi=1$ , (b) $\Omega_{max}/\xi=\Delta_{max}/\xi=2$, (c) $\Omega_{max}/\xi=\Delta_{max}/\xi=3$, (d) $\Omega_{max}/\xi=\Delta_{max}/\xi=4$. The inset in each subfigure highlights the corresponding time evolution of concurrence.}
    \label{fig:BBNNConcComp}
\end{figure}

\begin{figure}[!tbp]
  \begin{subfigure}[b]{0.4\textwidth}
    \includegraphics[width=\textwidth]{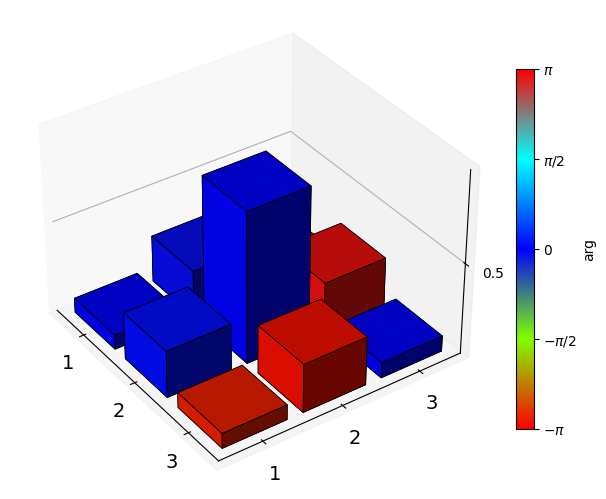}
    \caption{Density matrix histogram with $\Omega_{max}/\xi=\Delta_{max}/\xi=1$.}
    \label{fig:f1}
  \end{subfigure}
  \hfill
  \begin{subfigure}[b]{0.4\textwidth}
    \includegraphics[width=\textwidth]{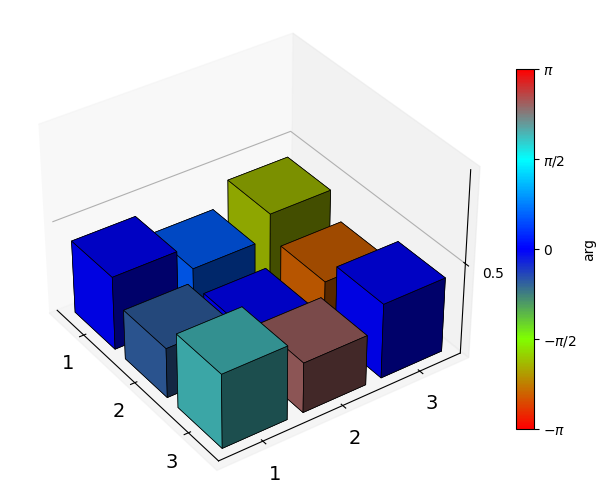}
    \caption{Density matrix histogram with $\Omega_{max}/\xi=\Delta_{max}/\xi=2$.}
    \label{fig:f2}
  \end{subfigure}
  \begin{subfigure}[b]{0.4\textwidth}
    \includegraphics[width=\textwidth]{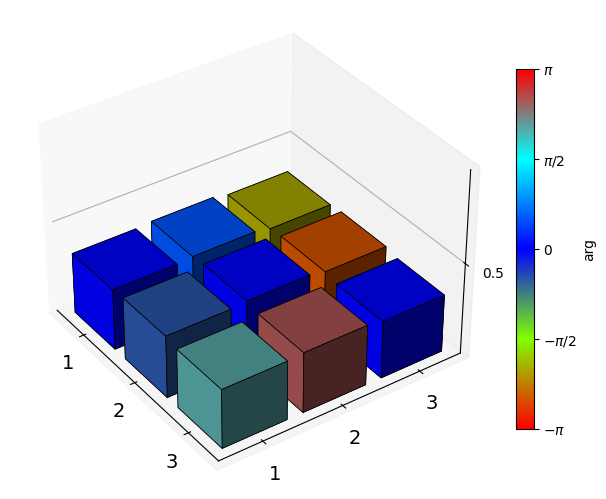}
    \caption{Density matrix histogram with $\Omega_{max}/\xi=\Delta_{max}/\xi=3$.}
    \label{fig:f3}
  \end{subfigure}
  \hfill
  \begin{subfigure}[b]{0.4\textwidth}
    \includegraphics[width=\textwidth]{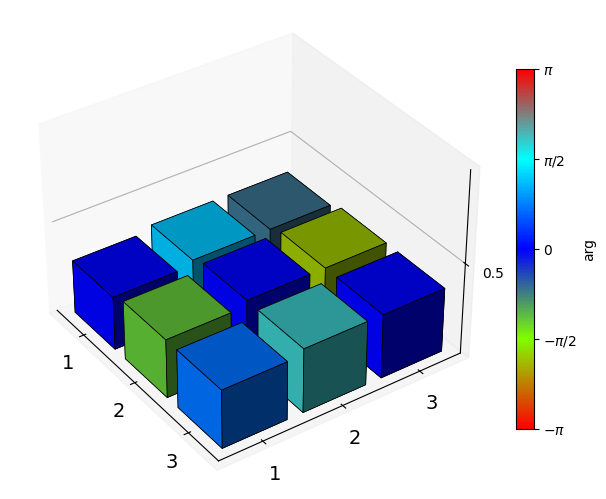}
    \caption{Density matrix histogram with $\Omega_{max}/\xi=\Delta_{max}/\xi=4$.}
    \label{fig:f4}
  \end{subfigure}
  \caption{Density matrix histograms on final states when optimizing concurrence with BBNN method.}
  \label{fig:histograms}
\end{figure}

\begin{figure}[H]
\centering
    \includegraphics[scale=0.4]{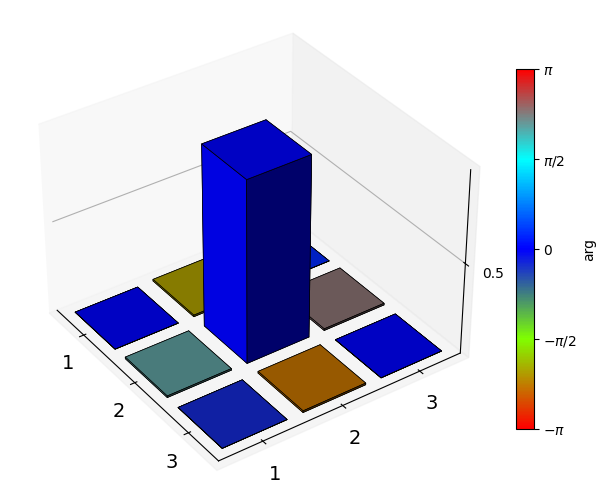}
    \caption{Density matrix histogram of final states when optimizing fidelity with BBNN method for $\Omega_{max}/\xi=\Delta_{max}/\xi=1$.}
    \label{fig:f5}
\end{figure}

As in the previous cases, in Figs. \ref{fig:BBNNFidDetailed} and \ref{fig:Tab:BBNNConDetailed} we present the optimal controls and the time evolution of populations and fidelity/concurrence, respectively, for control bound $\Omega_{max}/\xi=\Delta_{max}/\xi=1$. Notice once more that the desired concurrence level $\mathcal{C}>0.9999$ is obtained faster than the same Bell state fidelity level $\mathcal{F}>0.9999$, while the maximally entangled state reached is different than the particular Bell state, since the $P_2$ population in Fig. \ref{fig:Tab:BBNNConDetailed}(c) is noticeably less than unity. In Figs. \ref{fig:BBNNFidComp} and \ref{fig:BBNNConcComp} we show the optimal controls maximizing fidelity/concurrence (also displayed in the inset), respectively, for four different values of the control bound $\Omega_{max}/\xi=\Delta_{max}/\xi=1, 2, 3, 4$. As with the previous methods, in all cases the desired concurrence level is obtained faster than the targeted fidelity level, see Fig. \ref{fig:FinalTimingsPSO}(c), while a different maximally entangled state is reached, see Fig. \ref{fig:Simplex}(c). This is also illustrated in Fig. \ref{fig:histograms}, where we display the density operator corresponding to the state found when optimizing concurrence, for different maximum absolute control values, while in Fig. \ref{fig:f5} we display the density operator corresponding to the Bell state for comparison.
The significant changes in the optimal BBNN waveform observed in Fig. \ref{fig:BBNNConcComp} with the increase of the control bound, can be attributed to the discrete nature of the controls, while the PSO and SLSQP methods employ smooth parameterized control functions. The fine control discretization used in BBNN allows for large variations within small time intervals. As the control bound increases, the waveform is adjusted to drive the system at the corresponding closest point on the target manifold, and BBNN offers greater flexibility than the other methods, leading to the observed significant changes in the control shape.


\begin{figure}[H]
\centering
    \includegraphics[scale=0.4]{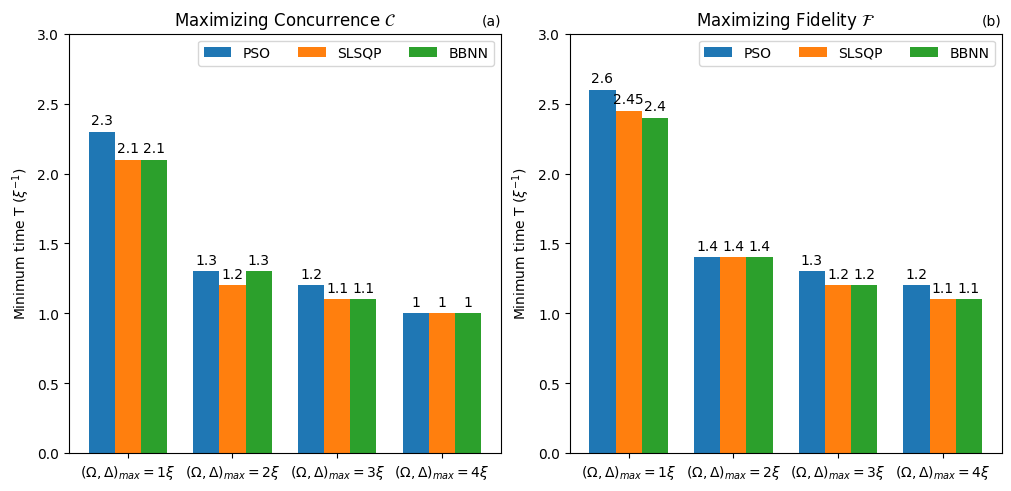}
    \caption{Comparison of the minimum time to reach the desired level of (a) concurrence $\mathcal{C}$ and (b) fidelity $\mathcal{F}$, with the PSO (blue), SLSQP (orange) and BBNN (green) methods, for different values of the maximum control amplitude.}
    \label{fig:Time_Comp}
\end{figure}

We close this section by comparing the three methods in terms of the minimum necessary time to reach the required concurrence and fidelity level, shown in Figs. \ref{fig:Time_Comp}(a) and \ref{fig:Time_Comp}(b), respectively, for the four values of the upper control bound used. We observe that in all cases the PSO method requires the longer times, while in most cases there is a tie between SLSQP and BBNN, except the occasion $\Omega_{max}/\xi=\Delta_{max}/\xi=2$ for concurrence where SLSQP is faster and the occasion  $\Omega_{max}/\xi=\Delta_{max}/\xi=1$ for fidelity where BBNN is faster. 

The better performance of SLSQP and BBNN methods compared to PSO in obtaining time optimal solutions which achieve the same desired accuracy ($0.9999$), may be explained by the nature of the methods since PSO is more probabilistic and does not use gradients. However, it is important to state that PSO can avoid local minimums and find the global minimum of an objective more easily than the other methods, a property that can be very beneficial in control landscapes with local minimums that are not easy to escape with a gradient-based optimization method. On the other hand, deep learning methods such as BBNN contain a lot of optimizable parameters with a lot of non-linearities introduced into the system, offering great potential for function approximation even without being physics-informed. As demonstrated by the results, BBNN can approximate the optimal control functions with great success and accuracy, strictly by learning from simulated data. The flexibility of the BBNN method is also illustrated in Fig. \ref{fig:BBNNConcComp}, where the optimal waveforms undergo significant changes as the control bound increases, to drive the system at the nearest maximally entangled state.

The complexity of the applied methods is different and not very comparable since their working and nature are quite different. The time complexity of the gradient-based methods BBNN and SLSQP is lower than the time complexity of PSO, due to the effectiveness of the gradient optimization methods and the probabilistic nature of PSO. However, BBNN and SLSQP experience higher memory and space complexity than the PSO, because of the need to keep track of the gradients during the training process. This could be a problem in previous decades, but current computational power performs this kind of calculation quite effectively.

All the software versions used in the present study are listed in Table \ref{software}, while the simulations were performed on a computer with specifications given in Table \ref{hardware}.

\begin{table}[ht]
    \begin{tabular}{ | c | c |} 
      \hline
      \textbf{Software} & \textbf{Version} \\
      \hline
      Pytorch & 2.3.0 \\ 
      \hline
      QuTiP & 5.0.1 \\ 
      \hline
      Python & 3.12.4 \\ 
      \hline
      Scipy & 1.13.0 \\ 
      \hline
    \end{tabular}
    \caption{Software~versions.}
    \label{software}
\end{table}

\begin{table}[ht]
    \begin{tabular}{ | c | c |} 
      \hline
      \textbf{Component} & \textbf{Model} \\
      \hline
      CPU & AMD Ryzen 5 5600X 6-Core Processor \\ 
      \hline
      Memory RAM & 32~GB \\ 
      \hline
    \end{tabular}
    \caption{Computer hardware~specifications.}
    \label{hardware}
\end{table}

\section{Conclusions}
\label{sect:concl}

We studied the problem of fast generation of concurrence in a system composed of two coupled spins, using the PSO and SLSQP algorithms as well as a DL method, to obtain the optimal controls (bounded time-dependent Rabi frequency and detuning) which drive the system in minimum time to a final state with concurrence level greater than $0.9999$. For the PSO and SLSQP methods the problem was formulated as a constrained optimization problem using a truncated trigonometric expansion of the controls with optimization parameters the coefficients of the harmonics. For the DL method, a simplified version of a PINN was used, called a BBNN, that only outputs the values of the control functions. The model is not directly physics informed, but it grasps the shapes of the controls needed to solve the problem. It is a very simple, yet quite efficient and accurate way to produce the optimal control functions. We also used these methods to find the optimal controls which drive the coupled spin pair to a specific Bell state, a problem which has been studied in previous works using other methods, like adiabatic rapid passage, shortcuts to adiabaticity and numerical optimal control, and for comparison reasons the target fidelity was also set to $0.9999$. We solved these problems using several different bounds on the controls. In all cases and for all methods we found that the targeted concurrence level is reached faster than the same fidelity level of the specific Bell state. As the control bound increases, the minimum time to reach the desired levels of concurrence/fidelity decreases. Additionally, for different control bounds, the corresponding optimal controls obtained for maximizing concurrence drive the system to a different final maximally entangled state.

The present work demonstrates that machine learning and optimization offer efficient techniques for the fast generation of entanglement in coupled spin systems, and we plan to extent it to systems involving more spins, for example chains. Also, the optimization of both fidelity and concurrence shows that the used methods are flexible enough to achieve multiple objectives within the same quantum setting with minor changes, only in the definition of the objective function. We close by pointing out that, although in the present paper we focus on studying the control problem numerically, we also plan to investigate it using optimal control theory following the steps of our recent work \cite{Evangelakos24}, where we derived optimal pulse-sequences for the fast charging of a quantum battery composed of a coupled spin system similar to the one considered here, with bounded global transverse field control.


\bmhead{Acknowledgements}
We acknowledge Prof. Ioannis Thanopulos for useful discussions during the development of this work. The publication fees of this manuscript have been financed by the Research Council of the University of Patras





\bmhead{Funding}
E.P. acknowledges support by the EU HORIZON-Project 101080085—QCFD. 

\bmhead{Conflict of interest/Competing interests}
Not applicable

\bmhead{Ethics approval and consent to participate}
Not applicable

\bmhead{Consent for publication}
Not applicable

\bmhead{Materials availability}
Not applicable

\bmhead{Code and data availability}
Code and data will be available in the public github repository https://github.com/dkoytrom/Fast-generation-of-entanglement-between-coupled-spins-using-optimization-and-deep-learning-methods.

\bmhead{Author contribution}
Conceptualization, D.S. and E.P.; methodology, D.K.; software, D.K.; validation, D.K., D.S. and E.P.; formal analysis, D.K. and D.S.; investigation, D.K.; data curation, D.K.; writing---original draft preparation, D.K, D.S. and E.P.; writing---review and editing, D.K., D.S. and E.P.; visualization, D.K. and D.S.; supervision, E.P. and D.S.; project administration, D.S. All authors have read and agreed to the published version of the manuscript.

\bmhead{Abbreviations}
The following abbreviations are used in this manuscript:\\

\noindent 
\begin{tabular}{@{}ll}
ML & Machine Learning \\
BBNN & Black Box Neural Network \\
DL & Deep Learning \\
PINN & Physics Informed Neural Network \\
PSO & Particle Swarm Optimization \\
SLSQP & Sequential Least Squares Quadratic Programming \\
SQP & Sequential Quadratic Programming 
\end{tabular}

\bigskip

\bibliography{sn-bibliography}

\end{document}